\documentclass{sig-alternate}

\usepackage{amsfonts,amsmath,amssymb} %% Additional math chars
\usepackage{graphicx}
\usepackage[utf8]{inputenc}
\usepackage[backend=bibtex, style=numeric, citestyle=numeric]{biblatex}
\usepackage[defblank]{paralist} % inline lists
\usepackage{url} 
\usepackage{specs}
\usepackage{subfigure}
\usepackage{xspace}
\usepackage{algorithm}
\usepackage{verbatim}

\usepackage{maths}
\usepackage{mymacros}

\usepackage{tikz}
\usepackage{import}
\usetikzlibrary{arrows,decorations.pathmorphing,backgrounds,
  fit,positioning,shapes.symbols,shapes,shapes.geometric, chains,shapes.multipart, calc}

\usepackage{listings}
\newboolean{showcomments}
\setboolean{showcomments}{false}
\ifthenelse{\boolean{showcomments}}
{ \newcommand{\mynote}[3]{
   \fbox{\bfseries\sffamily\scriptsize#1}
   {\small$\blacktriangleright$\textsf{\emph{\color{#3}{#2}}}$\blacktriangleleft$}}}
{ \newcommand{\mynote}[3]{}}

\usepackage{cleveref}

\begin{document}
%\setlength{\pdfpageheight}{\paperheight}
%\setlength{\pdfpagewidth}{\paperwidth}

%\conferenceinfo{SCRUM workshop, Middleware}{2015 Vancouver, Canada}
%\mainmatter

\title{Decoupling conflicts for configurable resolution in an open
  replication system}

%\author{Christian Weilbach \and Konrad K\"uhne \and Annette Bieniusa}

\numberofauthors{3} %  in this sample file, there are a *total*
\author{
% 1st. author
\alignauthor
Christian Weilbach\\ %\titlenote{Core developer of replikativ.}\\
       \affaddr{Universität Heidelberg, Germany} \\
       \email{christian@replikativ.io}
% 2nd. author
\alignauthor
Konrad K\"uhne\\ %\titlenote{Core developer of replikativ.}\\
       \affaddr{Universität Heidelberg, Germany} \\
       \email{konrad@replikativ.io}
% 3rd. author
\alignauthor
Annette Bieniusa\\ %\titlenote{Researcher in the SyncFree project funded by the EU.}\\
       \affaddr{TU Kaiserslautern, Germany}\\
       \email{bieniusa@cs.uni-kl.de}
}

\maketitle

%\listoftodos

%\todo{Conflict of interests statement?}

\begin{abstract}
  \texttt{replikativ} is a replication middleware supporting a new kind of
  confluent replicated datatype resembling a distributed version
  control system.  It retains the order of write operations at the
  trade-off of reduced availability with \emph{after-the-fact}
  conflict resolution. The system allows to develop applications with
  distributed state in a similar fashion as native applications with
  exclusive local state, while transparently exposing the necessary
  compromises in terms of the CAP theorem.  In this paper, we give a
  specification of the replicated datatype and discuss its usage in
  the \texttt{replikativ} middleware.  Experiments with the
  implementation show the feasibility of the concept as a foundation
  for \emph{replication as a service} (RaaS).
\end{abstract}

%A category including the fourth, optional field follows...
%\category{D.2.8}{Software Engineering}{Metrics}[complexity measures, performance measures]

%\terms{Demo}

\keywords{CRDT, eventual consistency, strong consistency, CAP theorem,
  middleware, replication, RaaS}

%\category{CR-number}{subcategory}{third-level}
\section{Introduction}

\begin{comment}
\begin{figure}
\centering
\includegraphics[width=0.7\textwidth]{./inconsistency.png}
\caption{Three replicas propagating an update each, which results in
  three different inconsistent values. Only specially crafted
  datatypes can converge under these conditions without a central
  agency coordinating sequential order of the operations.}
\label{fig:inconsistency}
\end{figure}
\end{comment}

While building scalable distributed systems, developers are typically
confronted with a number of (potentially conflicting) requirements
\footnote{A good summary from the developer of our browser-based
  database can be found at
  \url{http://tonsky.me/blog/the-web-after-tomorrow/}, or another
  similar perspective at
  \url{http://writings.quilt.org/2014/05/12/distributed-systems-and-the-end-of-the-api/}. Both
  links retrieved at 2015-07-18.}:
\label{requirements}
\begin{enumerate}
\item \textbf{Never lose data!}
\item \textbf{Always be available!} Even when
  offline.%, and be consistent after partitions.
\item Provide a simple API preferably with \emph{explicit consistency
  semantics}, e.g. \textit{DVCS-like}, and not a
  distributed file system without history, such as \texttt{dropbox}.
  Keep a sequential and consistent log of all modifications.
  % \todo{add copyright to dropbox, Riak, CouchDB, Datomic, git,...}
\item Support \emph{cross-platform} serialization while offering strong and
  extensible data semantics. Do not tie users to a single platform,
  e.g. \texttt{JSON} and \texttt{JavaScript}.
\item Replicate \emph{everything} at once consistently: code, data types and
  referenced binary values.
\item Do not require configuration of complex backend storage for simple
  applications.
  %Allow to implement totally \emph{decentral}
  %applications which do not rely on any particular infrastructure
  %provider and allow \emph{flexibility for redeployment} later.
\item Avoid ad-hoc reimplementation of network code with
  every application and framework.
  % \emph{complecting state with the
  %  network topology} of replicas and its failures.
\end{enumerate}

In our replication system, \texttt{replikativ},
\footnote{For a more detailed description have a look at the documentation at
  \url{https://github.com/replikativ/replikativ}.}
we combine a number of technologies to meet these requirements.
The main idea behind \texttt{replikativ} is to decouple the
replication of data from the application code, so that different
applications can share the same data base without mandatory agreement on
how the data is managed. This allows to fork application
state and innovate with new applications inside existent user bases
if the data is shared openly. Recent advances in machine learning
techniques, e.g. deep learning \cite{Bengio-et-al-2015-Book},
highlight that access to large amounts of data can unlock new
insights into different aspects of the involved
processes and allows to evolve smarter services. As of today,
the notion of open-source is often applied to code, while data is
considered as the property of individual
providers and only evaluated in their own interests. The greatest
potential of building shared data and knowledge bases is yet largely
untapped.

To provide users with sovereignty over their data, \texttt{replikativ}
will support a \emph{public-private key encryption} system and our
design already reflects that. It does not rely on the often false
security assumptions of a safe internal zone versus the internet and
instead will encrypt the data \emph{end-to-end} and not only the
communication channels.

In this paper, we focus on one technical core component of
\texttt{replikativ}. We designed the system around a new datatype, named
CDVCS, which decouples the conflict resolution mechanisms which
are typically hard wired in convergent replicated datatypes (CRDTs)
\cite{crdt_techreport11}. The contribution of this paper is the
documentation of this new datatype and its combination with different
conflict resolution strategies.
Originally inspired by \cite{histodb}, we use CDVCS to implement the
important concepts of a distributed version control system (DVCS) to
retain convergence and scalability in our replication system. We have
generalized \texttt{replikativ} furthermore to allow novel
combinations of CDVCS with arbitrary CRDTs by snapshot isolation. This
provides the developer a flexible design choices for different
trade-offs between consistency and scalability of write operations.

%\todo{The motivation is not clear yet. What do we want to sell here?}
% Power to the people (user/developer)? :) which motivates the
% decoupling of conflicts and gives a more flexible design space
% (either recovering other CRDTs or make supervised resolution?

\section{Related Work}

The design of CDVCS trades high \emph{availability} for weaker \emph{consistency} guarantees.
It is motivated by two major lines of work:
distributed version control systems (DVCSs) and confluent replicated data types (CRDTs).
For a general overview of consistency conditions and terminology, we refer to the survey of Dziuma et al. \cite{consistency13}.

\begin{comment}
\begin{figure}[h]
\centering
\includegraphics[width=0.3\textwidth]{./cap.png}
\label{fig:CAP}
\caption{While one cannot have consistency, partition-resistance and availability, many tradeoffs are possible. \cite{cap99}}
\end{figure}
\end{comment}

\subsection{DVCS}
Today's work flows for light-weight and open-source friendly software
development are centered around distributed version control systems
(DVCSs), such as \texttt{git}, \texttt{mercurial} or \texttt{darcs}.
Updates and modifications can be executed offline on the developer's
local replica(s) and are synchronized explicitly by her to the shared
code base.  In terms of the CAP theorem \cite{cap99},
\begin{comment}
seen in \Cref{fig:CAP}
\end{comment}
these systems provide availability, but allow divergence between
different replicas.  To reconcile the system state, some
\textit{after-the-fact} conflict resolution has to be applied,
e.g. through 3-way-merging mechanisms on text files or conflict
markers in case of non-mergeable differences.  These conflicts then
have to be resolved manually.  To support the user, DVCSs provide a
commit history which allows to determine the order of events and and
detect when consistency has been broken by concurrent writes.  While
this technique has proven very effective for source code, attempts to
transfer these systems to data have had limited success so
far. Prominent examples are file systems that have been built on top
of \texttt{git}, such as
\texttt{gitfs}\footnote{\url{https://github.com/presslabs/gitfs}} or
\texttt{git-annex}
\footnote{\url{https://git-annex.branchable.com/}}. There have also
been repeated attempts at using \texttt{git} directly to implement a
database.  Data management systems built on top of off-the-shelf DVCS
exhibit a number of problems:
\begin{enumerate}
\item Programs can exploit their text-oriented conflict resolution
  scheme by encoding the data in text format, e.g. in
  \texttt{JSON}. However, this requires serialization in
  \emph{line-based text-files} in a filesystem structure to be
  compatible with the default delta resolution mechanism for automatic
  conflict resolution. When the diff'ing of text files is customized
  in any of these DVCSs, usually a complete reimplementation of
  operations becomes necessary, and the desired compatibility is lost.
  Instead of relying on textual representation, we believe that
  providing customized data types with principled conflict resolution
  schemes is a more natural approach \cite{logoot}.

\item File systems are the \emph{historic} data storage model for a
  non-distributed \emph{low-level binary view} on data within a
  \emph{single} hierarchy (folders), and hence cannot capture and
  exploit higher-level structure of data to model and resolve
  conflicts.  Today, the preferred way to manage state from an
  application developer perspective is often a relational model or
  language-specific data structures as they are declarative and
  allow to focus on data instead of file system implementation details.

\item DVCSs often do not scale when it comes to handling of
  \emph{binary blobs} as they take part in the underlying delta
  calculation step. For example, \texttt{git} then needs an
  out-of-band replication mechanism like \texttt{git-annex} to
  compensate, adding additional complexity to the replication scheme.
\end{enumerate}

We think that these attempts based on DVCSs, while being close to our work,
are doomed to fail due to the trade-offs captured
by the CAP theorem. They try to generalize a highly optimized workflow
of a manual low frequency write-workload for development on source
code files to fast evolving high frequency write-workloads of
state transitions in databases. Much better trade-offs can be achieved
by picking the important properties of a DVCS and composing it with
other highly available data types.
% insert snapshot isolation?
This approach allows to build scalable,
write-workload oriented data types at the application level.

By building on the CDVCS, we can use other more efficient confluent datatypes for
\emph{write intensive} parts of the global state space, e.g. posts in
a social network and indexes on hashtags. A DVCS introduces
considerable overhead and potential loss of availability on these
operations.

\subsection{Confluent replicated datatypes (CRDTs)}
While the original motivation for our system was to implement a
DVCS-like repository system for an ACID database in an open and
partitioned environment of online and offline web clients and servers,
a replication mechanism was lacking.  DVCS systems like \texttt{git}
track only local branches and do not allow propagation of conflicts
and hence have no conflict-free replication protocol. Conflicts can
show up in any part of the network topology of replicas during
propagation of updates and they can only be resolved manually at this
position. Since the system has to stay available and needs to continue
to replicate at scale while being failure-resistant, we decided to
build on prior work on convergent replicated datatypes
\cite{crdt_techreport11}. CRDTs fulfill our requirements as they do
not allow and need any central coordination for replication. They also
provide a formalism to specify the operations on the datatype and
prove that the state of each replica always progresses along a
semi-lattice towards global convergence. CRDTs have found application
e.g. in \texttt{Riak}\footnote{\url{http://basho.com/tag/crdt/}}
%and a background presentation at \url{https://www.youtube.com/watch?v=1KP_pxFhlVU}}
or \texttt{soundcloud}\footnote{\url{https://github.com/soundcloud/roshi}}
%and an interview to the background at url{http://www.infoq.com/interviews/bourgon-crdt-go}}
to allow merging of the network state after arbitrary partitions without loss
of write operations. This is achieved by application of so called
\emph{downstream} operations on the respective local state of the CRDT. These
operations propagate as messages through the network. While
this fits our needs for the replication concept, it does not provide
semantics for strong consistency on sequential operations.

The notion of a CRDT in general implies automatic mergeability of
different replicas and does not lead to conflicts which then would
need some centralized information to be resolved. Hence, they are
usually referred to as \emph{conflict-free} replicated datatypes. Our
datatype somewhat breaks with this strong notion by merging conflicts,
emerging as branch heads, from the replication mechanism into the
value of the datatype. This allows resolution of conflicts at any
point in the future on any replica. CRDTs so far have mostly captured
operations on \emph{sets}, \emph{counters}, \emph{last-writer wins
  registers} (LWWR), connected \emph{graphs} and domain-specific
datatypes e.g. for \emph{text editing} \cite{crdt_techreport11}. None
of these datatypes allows to consistently order distributed writes.
Other CRDTs nonetheless have benefits
compared to our CDVCS datatype, because they cause less overhead on
replication and do not require conflict resolution with reduced
availability on application level, provided concurrency of the
datatype operations is acceptable. We hence generalized our
replication with a CRDT interface and reformulated our datatype
together with an OR-set in terms of this interface.

\begin{comment}
Similar comparable datatype concepts to CRDTs exist, there has been
for instance the development of \emph{cloud} \emph{datatypes}
\cite{cloudtypes12} which similarly to CRDTs try to raise the datatype
interaction level of commutative write operations to the
application.
\end{comment}
Similarly to CRDTs, cloud datatypes \cite{cloudtypes12} build on
commutativity of update operations.  The design still happens from a
cloud operator's perspective, though, as their \emph{flush} operation
allows \emph{explicit} synchronisation with some central view on the
data on a cloud server. All their non-synchronized datatypes can be
implemented with commutative CRDTs.

\begin{comment}
Their newer approach is made by using an operational model of a
replicated store called \emph{global sequence protocol}\cite{gdp_techreport_2015} which is
an adaptation of a total order broadcast. Each client and server
preserve consistency by tracking old and new operations in a sequence
and broadcasting for acknowledgement after new operations are received
while possibly seeing only a subsequence of the final sequence at any moment.
By using this methodology, the last writer according to the global sequence
always wins.
\end{comment}

Close to our work are versionable, branchable and mergeable datatypes
\cite{lorenz12}. This work models datatypes with an
object-oriented approach as a composition of CRDT-like
commutative datatype primitives (e.g. sets). To resolve conflicts, each
application needs to instantiate custom datatypes which
resolve conflicts at the application level.
%They demonstrate this with a hotel-booking system, which avoids
%overbooking. Similar to traditional CRDTs their datatypes require
%automatic conflict resolution during the replication against a central
%server of their counting process. Furthermore since each state is
%modeled as an application specific datatype,
Therefore, the code for conflict resolution has to be provided
consistently to each peer participating in replication. Having general
data types and compositions thereof in contrast allows us to replicate
without knowledge of the application and to upgrade the replication
software of the CRDTs more gradually, independent of application
release cycles. It also means that all peers can participate in the
replication no matter whether they have been assigned to an
application or not.

\texttt{swarm.js}
\footnote{\texttt{\url{https://github.com/gritzko/swarm}}} is the closest to our work.
It employs op-based
CRDTs for client replication and runs in the browser, allowing efficient offline
applications\footnote{\url{http://swarmjs.github.io/articles/2of5/}}.
In contrast, \texttt{replikativ} uses a dual representation of a CRDT,
  \emph{state-based} in-memory and \emph{op-based} on runtime during
  operations. This in-memory representation allows to store an
  efficient local compression of the operation history which is
  straightforward to implement for each CRDT and does not leak into
  the replication of operations.
  % In comparison \texttt{swarm.js} implements commutative datatypes and stores all operations.
  %Similar to our replication mechanism it does not require
  %the replication of the full history of operations on initial
  %connection, but only a separate state snapshot. It still needs to
  %store all operations for reconnections though.
  Further, \texttt{swarm.js} has not been designed as an open replication system.
  %for data exchange independent of a single cloud provider. This is
  %for efficiency reasons.
  It uses a spanning tree to minimize the
  replication latency of ops, while we build on a gossip-like protocol as
  building self-stabilizing spanning trees over the internet is still
  an open area of research \cite{DBLP:journals/corr/abs-0904-3087}.
  Our peer-protocol can be easily extended by middleware systems concerning
  just a single connection without dependencies on the code base.
  %\texttt{swarm.js} is limited to JavaScript and JSON, while our
  %system is host independent and also runs on the JVM with a port to
  %the CLR possible.  It has the powerful and extensible data semantics
  %of \texttt{edn}\footnote{\url{https://github.com/edn-format/edn}}.
  To our knowledge, \texttt{swarm.js} lacks a mechanism to exchange
  external values, most importantly (large) binary values. Our system
  uses referenced values by their platform independent hash, so
  datatypes only need to carry 32 bytes for every transaction. The
  referenced values need to be transmitted as well, of course, but can
  be structurally shared between datatypes and even commits.

\section{Application: Shared calendar}
\label{sec:application}

\texttt{replikativ} provides essential middleware functionality for implementing distributed applications,
covering client communication, durable data storage, and consistency management.
Let us sketch how application developers can employ this functionality.

\begin{figure}
\centering
\includegraphics[width=0.9\textwidth]{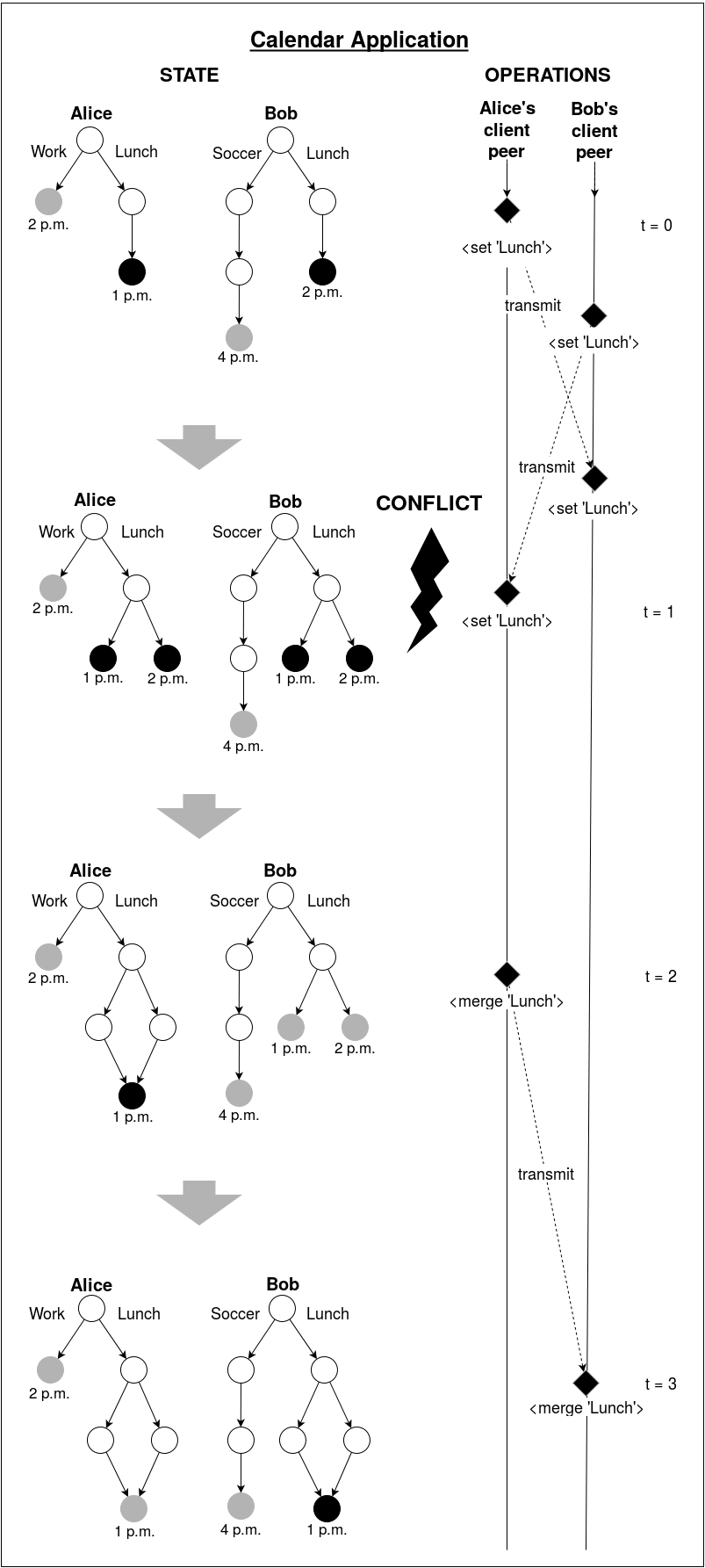}
\caption{States and operations side-by-side representing dynamic
  processes of an example calendar application where the CDVCS tracks
  the complete application state.  Each CDVCS has multiple branches
  that describe an appointment.  Both instances of the CDVCSs have a branch which is
  shared and one which is private to each CDVCS.  Conflicts can only
  arise in shared branches.}
\label{fig:application}
\end{figure}

As a basic example, consider a calendar application where people store
their appointments and synchronize them with others.
In the context of this paper, we simplify the calendar application by
tracking only titles of appointments and their time.  Each appointment
is tracked as a branch.
Let us assume that Alice and Bob want to use a shared calender
 to synchronize on a lunch appointment, alongside the otherwise private
appointment branches as shown in \Cref{fig:application}.  Alice has to
work at 2 pm, therefore she wants to eat lunch earlier at 1 pm. Bob
has soccer practice at 3 pm, so he prefers lunch actually later at 2 pm.
Once their clients are connected, both transmit their concurrent
operations to each other. This causes a conflict because they have set
a different time for their lunch. The application now notifies Alice and
Bob to resolve the conflict.  Alice merges both commits following a
user-moderated consistency scenario
\cref{sec:user-moderated-consistency}. The operation is then
transmitted to Bob's client and also applied there.

\section{CDVCS datatype}
\label{sec:model}

We compose a CRDT satisfying the requirements from
\Cref{requirements} by implementing a DVCS with the primitives
available from CRDTs. Our consistency requirement for an ACID
transaction log demands a sequential history. In DVCS, this is captured
by an \emph{add-only, monotonic} DAG of commits which represent
identities, i.e. values changing in time. The graph is monotonically growing
and can be readily implemented as a CRDT \cite{crdt_techreport11}.  To
track the identities in the branch, we need to point to their heads in
the graph.

\begin{comment}
as can be seen in \Cref{fig:application}.
\end{comment}

In a \textit{downstream} update operation to a branch with head $a$,
e.g. one reflecting a commit $b$, the branch heads are now
$\{a,b\}$. This is resolved in a DVCS by a \emph{lowest common ancestor
search (LCA)}. Whenever we want to resolve a branch value, i.e. its
history, we need to remove all stale ancestors and either have
only one head or a conflict of multiple ones. We therefore
remove stale ancestors in the set of heads on downstream operations,
so we do not need to use a CRDT for the branch heads.

 Combining the
DAG, the sets and LCA completes our CRDT which we refer to as
\emph{confluent} DVCS or \texttt{CDVCS}.

\paragraph{Correctness}
To show that \texttt{CDVCS} behaves properly as a CRDT, we have to
show that all operations satisfy the invariants of its metadata. In
particular the graph might never lose nodes or edges and always grow
according to the operation. All branch heads must always point to
leaves of the commit graph and might only be removed if they are
non-leaves (ancestors) of one of the others. For (operation-based) CRDTs, operations are
split in \emph{upstream} and \emph{downstream} operations, where the
former ones are applied at the local state of a replica, leaving the state unmodified,
and the latter ones are manipulating the state and are used to propagate the changes
also to the other replicas.

\paragraph{CRDT specification}
The correctness of CDVCS heavily relies on LCA which is used in a
typical DVCS to resolve conflicts. We use an online LCA version which
returns a set for common ancestors and the subgraphs traversed to
reach the ancestor(s) from each commit. We cover the following
operations, which we refer to by \emph{visited}:

\begin{figure}
\centering
\includegraphics[width=0.9\textwidth]{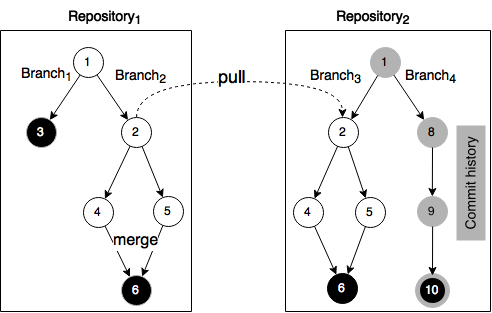}
\caption{The state of two repositories illustrating typical
  operations. Commits are represented as circles with black colored
  head commits. Both repositories have '1' as initial commit, one
  shared and one private branch each. While Branch2 exists before
  Branch3, Repository2 pulls all missing commits into a new branch
  from Repository1. Furthermore, by having two branch-heads in '4' and
  '5', a merge into '6' is applied based on some consistency
  scenario. The grey nodes represent the commit history from the head
  node '10' of Branch4 up to the root.}
\label{fig:operations}
\end{figure}

\begin{itemize}
\item \textbf{commithistory}:
Linearizes the history back to the root from some commit, e.g. head of
a branch, and loads all commit-values from memory as can be seen in
\Cref{fig:operations} on the right. It can be used to calculate the
current state of the application incrementally. It also covers the
branch history by providing the commit history of the current branch
head.
\item \textbf{commit}: Commits a new value to a branch. Since it just
  carries the edge and node added to the commit graph and the single
  new branch head in a set, the downstream operation will ensure that
  it is applied correctly.
\item \textbf{branch}: Creates a new branch given a parent. This
  operation forks off new branches directly at a commit without
  creating a new one. Since it just adds a new branch-id and initial
  head, the branch is correctly setup.
\item \textbf{pull}: Pulling adds all missing parent commits to the
  graph and adds the selected head into a set for the branch as can be
  seen in \Cref{fig:operations}.
\item \textbf{merge}: Merge resolves a conflict between multiple
  branch heads $H'$ by adding a new commit with $H'$ as parents as can
  be seen in \Cref{fig:operations} or \Cref{fig:application}.

\item \textbf{downstream} All operations only carry additions to the
  graph and sets of branch heads. We just have to apply all additions
  to the graph and merge the sets of heads. Since LCA properly detects
  all ancestoral heads, we can calculate the currently active heads
  safely by pairwise comparison. A special case is the initial
  full-state replication. Here the unknown part of the remote state is
  fetched and added to the own state by following the same procedure,
  which is also correct in this case. All dependencies are always
  fetched before atomic application, so the peer is in a
  self-consistent state and can act as a data provider for other peers
  if it is used with full replication.
\end{itemize}

\newnamex{commithistory}
\newnamex{commit}
\newnamex{pull}
\newnamex{mergebranches}

\begin{figure}[h]
 \begin{algorithmic}
   \Payload{graph $C$, set $H$}    \Comment{$C$: commit graph; $H$: branch heads}
   \Initial{$\{ r \rightarrow [] \}, \{ r \}$}
    \Query{\commithistory}{graph $C$, commit $c$}{$L$}
        \State{$S \gets emptystack()$}
        \State{$S.push(c)$}
        \Let{$L = \text{topological-sort}(C, [], S, \{\})$ %\Comment{See \Cref{alg:lin-hist}}
        }
    \EndQuery
    \Update{\commit}{commit $c$}
        \AtSource{$e$}
            \Let{$\tilde{C} = c \rightarrow [p]$}
            \Let{$\tilde{H} = \{c\}$}
            \EndAtSource
        \Downstream{$\tilde{C}, \tilde{H}$}
            \State{$C \gets C \union \tilde{C}$} %\Comment{Add the new commit edge and list of one vertex}
            \State{$H \gets removeancestors(H \cup \tilde{H})$  %\Comment{Compares pairwise and removes ancestors.}
            }
            \EndDownstream
        \EndUpdate
    \Update{\pull}{graph $\bar{C}$, commit $c$}
        \AtSource{$e$}
        \If{$\sharp H=1$}
            \Let{$h = H.pop()$}
            \Let{$\tilde{C} = lca(C,h,\bar{C},c).visited_c$}
            \Let{$\tilde{H} = \{c\}$}
        \EndIf
        \EndAtSource
        \Downstream{$\tilde{C}, \tilde{H}$}
            \State{$C \gets C \union \tilde{C}$}
            \State{$H \gets removeancestors(H \cup \tilde{H})$ }
            \EndDownstream
        \EndUpdate
    \Update{\mergebranches}{vector of commits $H'$}
        \AtSource{$e$}
            \Let{$\tilde{C} = e \rightarrow H'$}
            \Let{$\tilde{H} = \{e\}$}
        \EndAtSource
        \Downstream{$\tilde{C}, \tilde{H}$}
            \State{$C \gets C \union \tilde{C}$}
            \State{$H \gets removeancestors(H \cup \tilde{H})$ }
            \EndDownstream
        \EndUpdate

%    \LEL{$A$}{$B$}
%        \Let{$b = (lca(A,B).common = A.H)$}
    \MergeML{$S$} % \Comment{All downstream operations are (partial) state merge}
        \State{$C \gets C \union S.C$}
        \State{$H \gets removeancestors(H, S.H)$}
        \EndMergeML
 \end{algorithmic}
\caption{CDVCS: A DVCS-like datatype. Here we just model one branch to
  unclutter notation. Every DVCS with multiple branches can be represented by multiple clones of the same DVCS. $[...]$ denotes a vector of elements.}
\label{spec:DVCS}
\end{figure}

\section{Consistency scenarios}
Since the major difference of \texttt{CDVCS} compared to
commutative datatypes is the decoupled \emph{value-level} conflict
resolution, we now want to explore how this can be used to gain
different trade-offs between consistency and availability in
applications.

An important problem in distributed application design are the changing
scalability demands during the life cycle of an application. For
initial prototypes, no coordination or user moderated coordination
(\cref{sec:user-moderated-consistency}) might be sufficient. Once the
state space and workload increases, data moderated
(\cref{sec:data-moderated-consistency}) splits into commutative CRDTs
and CDVCSs can render the application both correct and efficient with
explicit semantics for the developer to monitor and optimize the data synchronization. A
relational query engine can be filled continuously with this mix of
datatypes decoupling the application level code from the CRDTs. If the
need for strong consistency arises, only some coordination
mechanism has to be added, while our replication protocol still takes
care of everything else. We pursue this strategy in our social network
demo application \texttt{topiq}\footnote{\url{https://topiq.es} The
  logic is executed client-side in the browser, the server only
  coordinates with pull-hooks.} with
\texttt{datascript}\footnote{\url{https://github.com/tonsky/datascript}}.

%AB: Mir ist der Sinn der ersten zwei subsections nicht klar.
%-> Wie verhalten die sich zum Use case? Kann man das in replikativ umsetzen? Falls nicht, braucht man es dringend?
\subsection{Strong consistency}
As an example for strong consistency, we consider the transaction log
of a typical \emph{ACID} relational database as is modelled in
\texttt{topiq}. Such a transaction log cannot be modeled by
traditional CRDTs in a system with distributed writes, since arbitrary
merges of non-commutative operations break consistency.

\begin{comment}
  as can be seen in \Cref{fig:inconsistency}
\end{comment}

\subsection{Single writer}
In a traditional database like
\texttt{Datomic}\footnote{\url{http://www.datomic.com/} A commercial
  scalable database with a relational query engine.} represented by
a linear transaction log, strong consistency can be modeled by having
a single writer with a single notion of time serializing the access to
the transaction log. CDVCS naturally covers this application case as a
baseline without conflict resolution.

\begin{comment}
  This is also the explicit design decision in
  \texttt{Datomic}\footnote{\url{http://www.datomic.com/}}, one
  inspiration for our work. In this case no branch conflicts can occur
  and our CRDT does not provide benefits. We can cover this scenario
  by allowing commit or \emph{non-conflicting} pull operations on a
  \emph{single} peer. Note that it might be internally distributed on
  a strongly consistent shared memory like Datomic, e.g. on a
  traditional database distributed in different
  data-centers. Modelling this with a branch in the DVCS is
  straightforward as it then can never be in a conflicting state.
  Note that it still allows to change the replication strategy for the
  CRDT later, so we can decouple the notion of strong consistency from
  the mechanism to achieve it.
\end{comment}

Interesting new choices are possible when different peers commit to
some branch creating different branch heads and the decoupled conflict
resolution comes into play. In these cases, conflicts can occur, but
they might still be resolvable due to application level constraints or
outside knowledge.

\subsection{User moderated consistency}
\label{sec:user-moderated-consistency}
In our replication system, each user can commit to the same
\texttt{CDVCS} on different peers at the same time only affecting her
own consistency. The user takes the position of the central agency
providing consistency. Consider as an example a private addressbook
application.  In this case, we can optimistically commit new entries
on all peers (i.e. mobile phone, tablet, notebook), but in case where
the user edits the same entry on an offline and later on an online
replica, a conflict will pop up once the offline replica goes back
online. Automatic resolution is infeasable because the integrity of
the entry without data loss can best be provided by the user. Since
these events are rare, user-driven conflict resolution is the best
choice and can be implemented by the application appropriately in a
completely decentralized fashion.

\subsection{Data moderated consistency}
\label{sec:data-moderated-consistency}
Similar to the hotel booking scenario in \cite{lorenz12}, we can allow
to book a room optimistically and then have \emph{one} DVCS in the
system updated strongly consistently on a peer which selectively pulls
and merges in all changes where no overbooking occurs. It provides a
globally consistent state and actively moves the datatype towards
convergence. The advantage of the \texttt{CDVCS} datatype is that this
decision can be done locally on one peer, independent of the
replication, while in \cite{lorenz12} the central peer needs to be
known and actively replicated to. Since the decision happens again in
a controlled, strongly consistent environment, it can happen
supervisedly and arbitrarily complex decision functions can be
executed atomically.  Assume for example that the preferences of a
user in a different CRDT or database allow rebooking rooms in a
comparable hotel nearby. In this scenario, the pulling operation can
decide to apply further transactions on the database to book rooms in
another hotel depending on information distributed elsewhere instead
of just rejecting the transaction. Furthermore, part of this
information could be privileged and outside of the replication system,
making it impossible in a system of open replication like ours to
automatically merge values on every peer. Conflicts in term of
\texttt{CDVCS} might in many cases still be resolvable by applying
domain knowledge.

\section{Evaluation}
%AB: What is a continuous evaluation?
%Maybe explain the context better: How many replicas/clients are using topiq.es? Since when?
%Average throughput?
We have continuously evaluated \texttt{replikativ} with \texttt{topiq}
on a diverse set of mobile and desktop browsers and found that the
replication behaves robustly despite the occasional inefficiencies
occuring during development. A second
application\footnote{\url{https://github.com/whilo/cnc}} is the
management of data from experiments run on a scientific simulation
cluster with the help of \texttt{Datomic}. In this case, the datatype
is used manually in an interactive REPL to track experiments including
results of large binary blobs, which is infeasible with \texttt{git}
or even a centralized \texttt{Datomic} alone.

%\subsection{Benchmark}

\begin{figure}
  \centering
\includegraphics[width=1.0\textwidth]{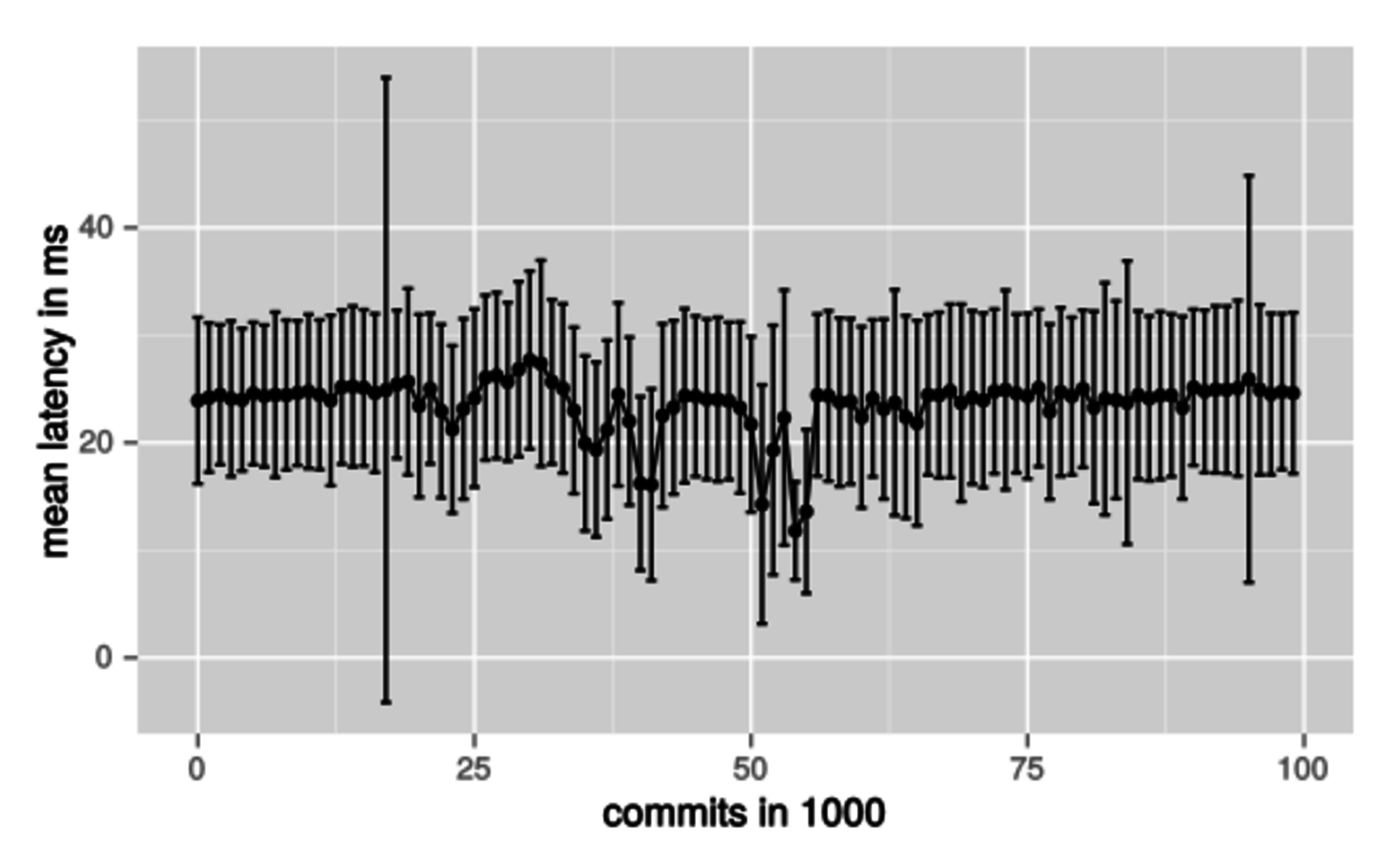}
\caption{Committing 100,000 times into a branch on one replica. This
  benchmark interacts with the full peer logic besides network IO.}
\label{fig:commitbench}
\end{figure}

Our work so far has mostly been focused on finding the proper
interfaces and levels of abstraction for \texttt{replikativ} to behave
correctly and reliably and allow straightforward optimized extension
to new CRDTs. But since performance and scaling of any distributed
system are critical and tradeoffs need to be known, we have conducted
some optimizations and run first benchmarks as you can see in
\Cref{fig:commitbench}. Most importantly commit times are hold almost
constant by application of \emph{Merkle-tree} like partitions of the
metadata.

\begin{comment}
The worst case runtime of the online LCA variant we use is
linear in the number of commits. To get an impression it can take a
dozen seconds to run LCA on a linear gap of a million commits on a 4
year old laptop. This can slow down the downstream replication logic
and increase latency when a deep subgraph has to be added. Several
preprocessing schemes for constant query time for LCA exist. But for
LCA on sparse DAGs the preprocessing time is so far at least of
quadratic complexity in the number of commits \cite{kowaluklca},
\cite{Bender05lowestcommon}. This renders them unusable for us. For
trees interesting sublinear online query variants exist
\footnote{\url{http://slideshare.net/ekmett/skewbinary-online-lowest-common-ancestor-search}},
but these make explicit use of the structural features of a tree and
cannot be transferred to DAGs. We decided to use a LRU caching
strategy and can reuse LCAs previously computed on subgraphs, so even
continuous conflicts in a branch will only need the LCA computation
once. Since the resolution of such large subgraphs will usually
involve a lot of network I/O to fetch the corresponding commit data,
it is not clear yet whether the LCA performance will be a bottleneck
for the usage of \texttt{CDVCS}.
\end{comment}

\section{Conclusion}

Together with our new datatype \texttt{replikativ} is a promising
platform to provide efficient \emph{replication} \emph{as} \emph{a} \emph{service}
(RaaS). Importantly, the available mix of datatypes  together with
\texttt{replikativ} allows to balance different
consistency vs. availability trade-offs depending on the application
semantics and scale. While we are now able to satisfy our initial
requirements, we are working on extended prototypes to benchmark and
verify our approach together with the open source community. As an
open and global network of replication, we plan to provide
support for application developers who do not want to care about
scaling of their backend either publicly or in private replication
networks. Already now, the development of the demo applications is
significantly easier than having a dedicated backend, and feels more
like management of local state in native applications than the typical
web development architectures. Cross-platform data semantics are
achievable. Since we explicitly build on the research around CRDTs, our
datatype semantics are transparent to the developer. Through the
implementation of new and modified CRDTs we will be able to adapt the
replication system to new needs while keeping old data and
applications available.

\begin{comment}
\section{Appendix}

\begin{algorithm}
\caption{DFS-like linearization of a commit history.}
\label{alg:lin-hist}
\begin{algorithmic}[1]
\Procedure{linearize-history}{graph $C$, seq $L$, stack $S$, set $V$}
\State $f \gets$ S.pop()
\State seq $ps \gets$ filter($\lambda.x$ $x \notin V$, parents(C,f))
\If{$f \not= nil$}
  \If{not empty($ps$)} \Comment{Keep recurring towards root.}
    \State $S$.push($ps$)
  \Else
    \If{$f \notin H$} \Comment{Only add commit on first occurance.}
      \State $L$.append($f$)
    \EndIf
    \State $H \gets H \cup \{ f \}$ \Comment{Remember the addition of f}
  \EndIf
  \State \textbf{return} \textsc{linearize-history}($C,L,S,V$)
\Else
  \State \textbf{return} $L$
\EndIf
\EndProcedure
\end{algorithmic}
\end{algorithm}
\end{comment}

%\bibliographystyle{splncs03}
%\bibliography{references}
\printbibliography

\end{document}